# The prevalence and impact of university affiliation discrepancies between four well-known bibliographic databases – Scopus, Web of Science, Dimensions, and Microsoft Academic

Philip J. Purnell
[1]Centre for Science and Technology Studies,
Leiden University,
P.O. Box 905, 2300 AX Leiden, The Netherlands
[2]United Arab Emirates University, Al Ain, UAE
Tel: +971 50 552 9356
p.j.purnell@cwts.leidenuniv.nl
ORCID: 0000-0003-3146-2737

# Abstract

Research managers benchmarking universities against international peers face the problem of affiliation disambiguation. Different databases have taken separate approaches to this problem and discrepancies exist between them. Bibliometric data sources typically conduct a disambiguation process that unifies variant institutional names and those of its sub-units so that researchers can then search all records from that institution using a single unified name. This study examined affiliation discrepancies between Scopus, Web of Science, Dimensions, and Microsoft Academic for 18 Arab universities over a five-year period. We confirmed that digital object identifiers (DOIs) are suitable for extracting comparable scholarly material across databases and quantified the affiliation discrepancies between them. A substantial share of records assigned to the selected universities in any one database were not assigned to the same university in another. The share of discrepancy was higher in the larger databases, Dimensions and Microsoft Academic. The smaller, more selective databases, Scopus and especially Web of Science tended to agree to a greater degree with affiliations in the other databases. Manual examination of affiliation discrepancies showed they were caused by a mixture of missing affiliations, unification differences, and assignation of records to the wrong institution.

# 1. Introduction

## 1.1. The problem of affiliations

The research community understands to varying degrees the importance of getting its university affiliation names right. Individual researchers are now routinely assessed at least in part on their ability to produce published articles and their institutions are at least partially ranked on those very same papers. The single factor tying the paper to its author's employer is the affiliation name given by the author when they submit the manuscript to a journal. There are many ways of acknowledging a university and one can easily confuse a rudimentary database by simply swapping My City University with University of My City. Other common variations involve acronyms, MCU, UMC, or partial acronyms, MC University, Univ MC. The list easily extends to dozens of variants when authors introduce their faculty or department name sometimes at the expense of the university name. Add to that the common practice of incorporating one institution into another or splitting part of a university away from its main organisation, along with larger mergers and creation of international branch campuses and we have a complex problem for those assessing the university's research output.

Indeed, nowadays journal and author names are relatively constant while it is not uncommon for university names to change. Although there have been several initiatives to address the problem by using unique identifiers for research institutions, none have been universally adopted to the same extent as for journals (ISSN), individuals (ORCID), or documents (DOI). These efforts, summarised in Table 1 have mainly been made by the major citation indexes such as Scopus (Affiliation Identifier or AFID), Web of Science (Organisation Enhanced), and Dimensions (Global Research Identifier Database or GRID). In addition, a new community-led collaboration of multiple organisations has been launched and is known as the Research Organisation Registry (ROR). It holds promise because it is closely linked to GRID and is to be incorporated into Crossref metadata (Lammey, 2020).

Table 1. Disambiguation processes used by bibliographic databases

| Database | Disambiguation | Abbr. | Process |
| --- | --- | --- | --- |
| Scopus | Affiliation identifier | AFID | Institutions assigned an 8-digit AF-ID and variants linked to main AFID |
| Web of Science | Organization enhanced | OE | Unifes the most frequently occurring address variants to preferred names |
| Dimensions | Global research identifier database | GRID | Freely accessible database of research organisations that are assigned a unique and persistent identifier linked to its variants. |
| Microsoft Academic | Global research identifier database | GRID | Freely accessible database of research organisations that are assigned a unique and persistent identifier linked to its variants. |
| Community led registry | Research organization registry | ROR | Community-led initiative to supersede GRID and be incorporated into Crossref metadata |

Databases used in bibliometric assessments have made strides into resolving the problem of disambiguation using different solutions including manual submission of affiliation variant lists by universities to database owners or automated unification systems. The degree of accuracy is still unquantified and policy makers who rely on bibliometric analysis often overlook an inherent level of error when using these data sources.

### 1.2. The importance of affiliations

Initiatives to identify and list the world's most influential academics based on citations to their work rely on affiliation disambiguation techniques. Clarivate's Highly Cited Researchers list recognises approximately 6,000 scientists who have published papers cited in the top 1% of their field in the preceding decade. The list is published once a year and is searchable by academic institution. The composition and validity of the list is therefore dependent on the affiliation disambiguation performed by Clarivate on its underlying Web of Science database. Similarly, the recently updated 'Stanford' author database of standardised citation metrics (Ioannidis, Boyack, & Baas, 2020) relies on Scopus affiliation disambiguation. The list of top 2% authors can be downloaded and is sortable by academic institution. As universities seek to recruit scientists who appear on these prestigious lists, an academic's value is increased by virtue of their presence. The ability of the databases to accurately link authors to their affiliations is therefore increasingly valued by recruitment professionals as well as research managers.

Many universities driven by increased external competition (Brankovic, Ringel, & Werron, 2018; Espeland & Sauder, 2016) seek to maximise their position in the various international ranking tables. The ranking organisations in turn typically assess institutions' performance against a set of criteria that usually include the quantity and impact of research publications (Centre for Science & Technology Studies Leiden University, 2020; QS Intelligence Unit, 2019; Shanghai Ranking Consultancy, 2019; Times Higher Education, 2021; US News & World Report LP, 2019). These ranking systems use either Elsevier's Scopus or Clarivate's Web of Science (Web of Science) to compute the bibliometric component of their tables. Nature Index, a database of author affiliations and institutional relationships also recently used data from Dimensions in an experiment to identify research in the field of Artificial Intelligence (Armitage & Kaindl, 2020).

The level of accuracy of those databases and their ability to assign papers to the correct affiliations consequently becomes one of the limiting factors in an institution's performance (Orduna-Malea, Aytac, & Tran, 2019). Any 'missing' papers can cost a university valuable places in the ranking table and authors are routinely encouraged to use the official institution name when publishing. Nevertheless, each year universities complain to the ranking systems of missing papers, but the rankers are constrained by the limitations of the citation indexes.

Disgruntled universities are usually referred to the database owners to resolve their affiliation-related complaints.

Times Higher Education routinely uses the Scopus affiliation as delivered by Elsevier although it has occasionally worked directly with institutions to ensure the mapping used in the ranking coincides with their organisational structure. QS receives data from Scopus and then groups distinct Scopus AFIDs including medical schools, business schools, hospitals, and technical research institutes into single university entities that match those in its rankings database. In this process, QS relies on input from the institutions to define such relationships (QSIU, 2019). The Leiden Ranking uses its own (CWTS) version of the Web of Science and conducts additional rounds of affiliation disambiguation. Specifically, the CWTS system unifies all address variants that occur at least five times in the Web of Science database, identifies missing university affiliations from departments and city names, attributes papers from hospitals and medical centres to their affiliated universities based on author publication rules (Centre for Science & Technology Studies Leiden University, 2020; Van Raan, 2005; Waltman et al., 2012).

Comparison of database coverage has become easier since most now aim to link their records to the Digital Object Identifier (DOI). The DOI has become established as a persistent, reliable identifier together with a system using that identifier to locate digital services associated with that content (e.g. Gasparyan, Yessirkepov, Voronov, Maksaev, & Kitas, 2021; Zahedi, Costas, & Wouters, 2017). In bibliometric studies, the DOI can therefore be used as a common identifier to determine overlapping coverage of records in different databases. The existence of a DOI for a research article depends on the publisher generating the DOI and linking it to the article in the Crossref database. Most publishers now aim to do this routinely but there are plenty of records that do not have a DOI and that limits our ability to use it as a common identifier. Factors associated with lower prevalence of DOIs include non-academic records, arts and humanities fields, document types from books and conference proceedings. This study used DOIs to retrieve papers from selected universities and it was important as a first step to understand what proportion of the actual output we were looking at.

### 1.3. Research design

This paper addresses affiliation disambiguation by attempting to answer the following questions:

1. To what extent do discrepancies in author affiliations exist between the major bibliographic databases? Which databases have most discrepancies?

2. What types of discrepancies can be identified?

3. Are different types of discrepancy more prevalent in different databases?

The answers to these questions will be useful in our understanding of the extent to which research outputs are accurately assigned to universities by the different databases. Since many decisions are based on the outcome of bibliometric studies, comparisons, and university rankings, policy makers will be better informed about the limitations of bibliometrics studies and comparisons. The ranking bodies may take these limitations into account when they publish their league tables. Database owners may incorporate these findings into their development plans and algorithms to improve the accuracy of their products and make them more competitive.

We selected 18 universities for the study, all from the Arab region. Local or regional languages have been found to contribute to mistakes in author affiliations (Bador & Lafouge, 2005; Falahati Qadimi Fumani, Goltaji, & Parto, 2013; Konur, 2013) and to our knowledge, no such study of university affiliations in the Arab region has been published. We selected universities from Gulf countries, the Levant, and North-East Africa because of our familiarity with the region, language, and institutions.

We used a recent five-year time window (2014-18) and extracted records from four major databases, Scopus, Web of Science, Dimensions, and Microsoft Academic.

The first part of the study was to determine the proportion of publications from our selected universities that had DOIs. Records were extracted using DOI and compared with the total number of records for the selected publication years. This was to confirm we could use the DOI to identify a sufficient quantity of scholarly material.

Each database has taken a different approach to affiliation disambiguation, and it is important to know how they compare. Therefore, in the second part of the study we quantified the affiliation discrepancies between databases. We paired the databases and specifically looked at the non-overlapping records i.e., the surplus of records that were retrieved from one database in the pair but not the other. Records could be in the surplus because of discrepancies in affiliation or publication year, or differences in database coverage. We quantified the surplus in each database with respect to the other three for all the universities. We then calculated the proportion of the surplus that was caused by each of the three reasons (affiliation discrepancy, publication year discrepancy, or coverage differences). Discrepancies due to publication year were negligible and grouped together with coverage differences.

The third part of the study concentrated exclusively on those records that were caused by discrepancies in the university affiliation. We manually examined two dozen records from each database pair surplus and attempted to explain the discrepancies in the context of affiliation indexing. This has major implications for any university benchmarking study and particularly the international university rankings. The final section of the study requires close knowledge of university names and we therefore chose to make this a regional study.

## 2. Literature review

Work on author affiliations began in the mid-1980s. For example, the LISBON Institute, the predecessor of CWTS already used affiliation data from the Science Citation Index (SCI) to report the changes in academic collaboration in the Arab region following the geopolitical developments in the 1980s (DeBruin, Braam, & Moed, 1991). The problems of author affiliations became more important as bibliometric reports gained popularity and started being offered as a service (Calero-Medina, Noyons, Visser, & De Bruin, 2020). As in-house citation indexes were developed in the 1990s, care was taken to construct them in a manner that facilitated affiliation disambiguation. The problem of missing author affiliations has been largely addressed but in 2015 the Web of Science indexes still contained sizeable quantities of publications without any affiliations whatsoever (SCIE: 7.6%, SSCI: 6%, and A&HCI: 35%) (Liu, Hu, & Tang, 2018).

The owner of Web of Science tried to overcome the affiliation disambiguation problem by introducing its organisation enhanced feature. The disambiguation process involves creating normalised address segments for each record and unifying the most frequently occurring address variants to a list of currently over 14,000 preferred names (Clarivate, 2020a). Smaller or less frequently occurring organisations might not have a preferred name in the system or some of its legitimate address variants might not have been unified. Full details of the process are not available, but one imagines it a significant task to perform and then keep up to date in the light of organisation mergers and divestments. Organisations may request unification or corrections to its address variant list via an online form (Clarivate, 2020b).

The organisation enhanced feature has been selectively used in bibliometric studies to improve accuracy (Baudoin, Akiki, Magnan, & Devos, 2018). A recent study (Donner, Rimmert, & van Eck, 2020) showed widely varying recall and precision across institutions between the Web of Science organisation enhanced feature, Scopus AFID and a German institution affiliation disambiguation system described as 'near-complete' for German public research organisations. Taking the German system as ground truth, neither Web of Science organisation enhanced nor Scopus AFID provided high recall rates, and both showed widely varying precision across institutions impacting bibliometric indicators. The authors concluded the resulting inconsistencies in publication and citation indicators using the commercial vendor systems should be taken into consideration by policy makers.

A large-scale study comparing database coverage across a multi-institution dataset between Scopus, Web of Science, and Microsoft Academic examined the publication overlap of 15 universities with DOIs serving as the common attribute (Huang et al., 2020). The authors created a Venn diagram for each university showing the proportion of DOIs indexed by all three data sources. The diagram also revealed the extent to which DOIs were covered by only one of the three databases, which we term a surplus. For instance, DOIs found in Web of Science but not in Scopus or Microsoft Academic count as Web of Science surplus. The authors found that Microsoft Academic had the broadest DOI coverage but least complete affiliation metadata. The authors concluded that assessment of any institutional performance will produce different results depending on the source database. They went on to demonstrate that representation in databases varied widely between institutions, which compounds the likelihood of inaccuracies when comparing universities with each other. This introduces potential bias in any bibliometric assessment such as university ranking that relies on a single source. The present study builds on the work by Huang et al. (2020) by further analysing the affiliation assignation of DOIs across the same three databases plus Dimensions for 18 universities in a specific geographical region.

Another large scale study (Visser, Eck, & Waltman, 2021) compared coverage and the completeness and accuracy of citation links between five databases, namely Scopus, Web of Science, Dimensions, Microsoft Academic, and Crossref. The authors used a thorough and complex matching technique that compared pairs of records against seven bibliographic attributes and computed a matching score. They found Microsoft Academic had by far the broadest coverage of the scientific literature. In pairwise analysis, Scopus contained large quantities of exclusive publications not indexed in the other databases. Meanwhile, Dimensions/Crossref and Microsoft Academic indexed substantial bodies of records not found in Scopus. The Web of Science surplus compared with Scopus was smaller and limited mainly to meeting abstracts and book reviews which are not indexed in Scopus. Another important finding was that Dimensions and Crossref had almost overlapping coverage, which can be attributed to Dimensions' reliance on Crossref as its core content.

A comparative analysis at institutional level found more institutions had greater coverage in Scopus than in Dimensions, and that up to half the documents indexed in Dimensions have no institutional affiliation (Guerrero-Bote, Chinchilla-Rodríguez, Mendoza, & de Moya-Anegón, 2021). In the view of the authors, this invalidates Dimensions as a suitable data source for assessing university impact.

## 2.1. Limitations of DOI accuracy

All studies using DOI are susceptible to a number of limitations. First, DOI assignation is not always accurate. Indeed, one group found errors in 38% of the DOIs in the cited references of their sample from Web of Science (Xu, Hao, An, Zhai, & Pang, 2019). Most (92%) of the errors in the DOI were in the prefix and often included a surplus 'DOI' or a duplication of the DOI. The authors went on to propose an algorithm for cleaning the DOIs in the cited reference database. Another study found 8,841 'illegal' DOIs (defined as those that do not begin with '10') in Scopus (M. Huang & Liu, 2019) and referred to Elsevier's efforts to clean the Scopus data.

(Zhu, Hu, & Liu, 2019) created a search string to identify DOIs with errors in Web of Science. They wrote a search strategy aimed at identifying cases in which a numeric digit such as '0' had been replaced with the letter 'o'. Similar errors occurred where the number '6' had been confused with the letter 'b', or the number '1' with the letter 'l' among many other examples identified. In some of these cases, searching both the correct version of the DOI (with the numeric '0') or the erroneous version (with the letter 'o') returned the Web of Science record. Another problem is duplicate DOIs, this is when a single DOI is linked to multiple papers as reported by Franceschini, Maisano, & Mastrogiacomo (2015) or to multiple versions of the same publication (Valderrama-Zurián, Aguilar-Moya, Melero-Fuentes, & Aleixandre-Benavent, 2015). Databases have taken different approaches to this problem and Elsevier has recently invested in improving the Scopus data completeness and accuracy (Baas, Schotten, Plume, Côté, & Karimi, 2020).

The DOI was introduced in 2000 although there have been a number of other unique identifiers for published research papers. With the advent of electronic publishing, the Uniform Resource Locator (URL) was an early candidate. A study of more than 10,000 MEDLINE abstracts in 2008 looked at the presence of decay rate of URLs between 1994 and 2006 (Ducut, Liu, & Fontelo, 2008). The results showed that most (81%) of the URLs were available but that only 78% of the available URLs contained the actual information mentioned in the MEDLINE record, and one in six (16%) of the total were "dead" URLs. A study comparing multiple identifier systems found the DOI to be among the best following evaluation against seven criteria including identifier features, digital coverage, and comprehensiveness of scope (Khedmatgozar & Alipour Hafezi, 2015).

Over time, publishers incorporated DOIs into their online metadata. An examination of Web of Science and Scopus records revealed that by 2014, most (90%) of 'citable items' defined as journal articles, reviews and conference proceedings papers were being assigned DOIs in the sciences and social sciences (Gorraiz, Melero-Fuentes, Gumpenberger, & Valderrama-Zurián, 2016). The figures were lower for all document types and much lower in the arts & humanities, 50% for journal citable items and just 20% for books and book chapters. Articles published in regional publications may also be less associated with DOIs, a sample of scholars from Brazil showed DOIs among less than half of their journal papers and less than a tenth of their conference papers (Rubim & Braganholo, 2017). Meanwhile, Mugnaini et al. (2021) found the presence of international co-authors increased the proportion of DOIs.

## 3. Data and methods

This study examines the overlap of indexed DOIs for 18 universities (Table 1) of all document types, in four international, multidisciplinary bibliometric databases often used in bibliometric studies, namely Scopus, Web of Science, Dimensions, and Microsoft Academic.

From the Web of Science, we used five citation indexes; the Science Citation Index Expanded, Social Sciences Citation Index, Arts & Humanities Citation Index, the Conference Proceedings Citation Index-Science, and the Conference Proceedings Citation Index-Social Sciences & Humanities. We used neither the Book Citation Index, nor the Emerging Sources Citation Index because we do not have access to them. From Dimensions, we extracted only publications because these are comparable with the documents in the other databases, but not grants, patents, datasets, clinical trials, or policy documents. All data were retrieved from the Centre for Science and Technology Studies (CWTS) database system at Leiden University. The data was received in March 2020 (Web of Science), April 2020 (Dimensions), July 2020 (Microsoft Academic), and April 2021 (Scopus).

We extracted records based on the highest level of affiliation disambiguation available for the selected universities in each database. In Scopus, we used the Affiliation identifier (AFID) which is a unique identifier for the institution to which records are tagged. In many cases, Scopus includes one AFID for the overall organisation and multiple additional AFIDs for partner institutions and component units to reflect the organisational structure of the university. This enables the user to search for records from the entire organisation or for its sub-units individually. This process appears to have been conducted more rigorously for some universities than others. The result is that for some universities there is only one AFID while for others there are many and we therefore used only the main AFID for each university. From Web of Science, we used the organisation enhanced feature which is a preferred institutional name searchable in the database and to which records from that organisation and its component parts are unified. The organisation enhanced unification process is performed by the database owner with voluntary input from the institutions themselves and suffers from the same inconsistencies as the Scopus AFIDs. Therefore, we used only the high-level organisation enhanced name and no additional variants. Dimensions and Microsoft Academic each use GRID which was developed by Digital Science to describe both parent-child relationships between institutions and external related organisations. GRID disambiguates affiliation names for approximately 100,000 organisations and we therefore used the GRID record linked to the generally accepted name for each of the 18 university names.

Table 2. The universities used along with their abbreviations

| Abbreviated name | Full institutional name | Country |
|---|---|---|
| Ain Shams | Ain Shams University | Egypt |
| Alexandria | Alexandria University | Egypt |
| Assiut | Assiut University | Egypt |
| AUB | American University of Beirut | Lebanon |
| Babylon | University of Babylon | Iraq |
| Baghdad | University of Baghdad | Iraq |
| Bahrain | University of Bahrain | Bahrain |
| Carthage | University of Carthage | Tunisia |
| Jordan | University of Jordan | Jordan |
| Khalifa | Khalifa University | United Arab Emirates |
| King Abdulaziz | King Abdulaziz University | Saudi Arabia |
| King Saud | King Saud University | Saudi Arabia |
| Kuwait | Kuwait University | Kuwait |
| Qatar | Qatar University | Qatar |
| Lebanese | Lebanese University | Lebanon |
| Sfax | University of Sfax | Tunisia |
| Sultan Qaboos | Sultan Qaboos University | Oman |
| UAEU | United Arab Emirates University | United Arab Emirates |

In the first part of the study, we determined the proportion of records with a DOI in each database. We extracted the unique DOIs present in each database between the years 2014 and 2018 inclusive and calculated the proportion of total records in the database with a DOI. The overall share is reported as well as a share for each of the studied universities.

The second section of the study quantifies the share of the surplus caused by affiliation discrepancy and separately, the share of the surplus caused by database coverage. For this we used only those records with DOIs in the same five-year period and for the 18 selected universities. Any records whose publication year differed between databases meaning they were included in the five-year window in one database but not another, were counted under coverage discrepancy. Preparatory work for the study demonstrated the number of records in this category was negligible. Because of the complexities of database comparison, we compared coverage between the databases in pairs. That is to say we extracted DOIs using university affiliation names and the publication time window in one of the databases ('Primary'). We then searched for those same DOIs in the second database ('Comparator'), restricting the search to the same affiliation and the same publication time window. Those DOIs found in both databases constitute overlapping coverage and were not studied further. Those DOIs not found in the comparator database are termed the 'surplus' and the reasons for their absence are investigated. To establish these reasons, we systematically repeated the comparison removing elements of the search in the comparator database for affiliation and publication time period.

In the final part of the study, we manually analysed a sample of 24 affiliation discrepancies for each of the 12 database pairs. Examples were selected at random with a maximum three publications in each comparison from the same publisher. These examples served to illustrate the presence and types of affiliation discrepancies between databases. For each database pair surplus, we selected the university with the highest proportion of affiliation discrepancies. To have as much diversity as possible in the universities studied, each university was selected at most once. (If the university with the highest proportion of affiliation discrepancies had already been selected for another database pair, we moved on to the university with the next highest proportion.) Examination was performed by manually searching for the records on the web interface versions of each database and checking the published PDF documents as the ground truth.

# 4. Results and discussion

## 4.1. Number and proportion of DOIs

The total number of records along with the number of unique DOIs retrieved for each of the four databases is shown in Table 2.

Scopus and Web of Science each had DOIs for more than three quarters of the publications. Both these databases incorporate a selective expert assessment of journals, books, and conference series before they are admitted to the database. This process ensures that most indexed articles go through the established scholarly publishing route and are therefore increasingly likely to have a DOI. The proportion of records with DOIs for the selected universities in this study (89.9% Scopus and 82.3% Web of Science records) seem somewhat higher than the proportions reported by Gorraiz et al. (2016). However, that study used earlier data up to 2014 and demonstrated an upward trajectory for documents with a DOI that would roughly coincide with the figures presented here. Records with a DOI comprise 96% of the Dimensions database, which is the highest share among the four. As Dimensions uses Crossref as the key pillar of its content along with PubMed, the high proportion is not especially surprising. All Crossref records have a DOI and 90% of PubMed articles already had DOIs by 2015 (Boudry & Chartron, 2017).

There is some variance in DOI prevalence between the universities. The variance might be due to the over representation of certain document types that are less frequently assigned DOIs. e.g., books, book chapters, and conference proceedings. Equally, universities that publish more papers in subject fields such as the arts and humanities could also contribute to a lower proportion of DOIs among their publications (Gorraiz et al., 2016).

For studies concerning detailed comparison of database coverage, additional metadata such as article title and author names should be used to maximise identification of overlapping records. The focus of the current study however was not coverage, but the prevalence of affiliation discrepancies between databases. Therefore, we could exclude records that do not have DOIs without negatively impacting the study results. It was more important to identify records with a common identifier so we could compare their affiliations in the different databases. This approach was largely inspired by the work of Huang et al. (2020) who also took this approach.

Table 3. Number and proportion of records with DOI by university

| Institution | Scopus | | | Web of Science | | | Dimensions | | | Microsoft Academic | | |
|---|---|---|---|---|---|---|---|---|---|---|---|---|
| | DOI | Total | % DOI | DOI | Total | % DOI | DOI | Dimensions Total | % DOI | DOI | Total | % DOI |
| Ain Shams | 7,785 | 8,567 | 90.9% | 6,621 | 8,139 | 81.3% | 9,741 | 9,809 | 99.3% | 10,834 | 11,230 | 96.5% |
| Alexandria | 6,875 | 7,524 | 91.4% | 5,277 | 6,584 | 80.1% | 7,218 | 7,280 | 99.1% | 7,198 | 7,528 | 95.6% |
| Assiut | 4,608 | 5,027 | 91.7% | 3,930 | 4,762 | 82.5% | 4,439 | 4,511 | 98.4% | 4,538 | 4,829 | 94.0% |
| AUB | 4,364 | 4,733 | 92.2% | 4,420 | 5,569 | 79.4% | 4,080 | 4,115 | 99.1% | 5,620 | 5,961 | 94.3% |
| Babylon | 1,269 | 1,978 | 64.2% | 389 | 472 | 82.4% | 618 | 621 | 99.5% | 701 | 878 | 79.8% |
| Baghdad | 2,687 | 3,827 | 70.2% | 1,380 | 1,629 | 84.7% | 1,930 | 1,959 | 98.5% | 2,214 | 3,003 | 73.7% |
| Bahrain | 811 | 952 | 85.2% | 506 | 566 | 89.4% | 722 | 723 | 99.9% | 722 | 778 | 92.8% |
| Carthage | 4,595 | 4,957 | 92.7% | 3,985 | 5,263 | 75.7% | 4,103 | 4,113 | 99.8% | 3,257 | 3,433 | 94.9% |
| Jordan | 3,725 | 4,334 | 85.9% | 2,599 | 3,132 | 83.0% | 3,277 | 3,306 | 99.1% | 3,758 | 4,441 | 84.6% |
| Khalifa | 5,781 | 6,252 | 92.5% | 3,936 | 5,176 | 76.0% | 4,991 | 4,991 | 100.0% | 3,746 | 4,031 | 92.9% |
| King Abdulaziz | 22,038 | 24,045 | 91.7% | 19,287 | 21,502 | 89.7% | 19,080 | 19,221 | 99.3% | 18,966 | 19,889 | 95.4% |
| King Saud | 21,620 | 23,870 | 90.6% | 18,246 | 21,457 | 85.0% | 19,470 | 19,788 | 98.4% | 19,165 | 20,305 | 94.4% |
| Kuwait | 3,204 | 3,617 | 88.6% | 2,406 | 2,976 | 80.8% | 3,002 | 3,021 | 99.4% | 3,040 | 3,280 | 92.7% |
| Lebanese | 2,880 | 3,107 | 92.7% | 1,590 | 2,198 | 72.3% | 2,933 | 2,967 | 98.9% | 2,884 | 3,148 | 91.6% |
| Qatar | 6,080 | 6,686 | 90.9% | 4,182 | 5,393 | 77.5% | 5,888 | 5,895 | 99.9% | 5,980 | 6,373 | 93.8% |
| Sfax | 7,339 | 7,863 | 93.3% | 6,296 | 8,155 | 77.2% | 6,793 | 6,803 | 99.9% | 6,055 | 6,340 | 95.5% |
| Sultan Qaboos | 3,727 | 4,229 | 88.1% | 2,650 | 3,241 | 81.8% | 3,485 | 3,539 | 98.5% | 4,471 | 4,898 | 91.3% |
| UAEU | 3,699 | 4,090 | 90.4% | 2,540 | 3,172 | 80.1% | 3,393 | 3,419 | 99.2% | 3,327 | 3,626 | 91.8% |
| Total 18 universities | 107,578 | 119,720 | 89.9% | 85,069 | 103,363 | 82.3% | 100,511 | 101,390 | 99.1% | 102,145 | 109,503 | 93.3% |
| Whole database | 13,046,237 | 15,495,969 | 84.2% | 9,833,766 | 12,975,857 | 75.8% | 21,691,146 | 22,589,839 | 96.0% | 21,042,708 | 55,168,811 | 38.1% |

## 4.2. Prevalence of affiliation discrepancies

In the next part of the study, we examined the prevalence of affiliation discrepancies in four citation indexes, Scopus, Web of Science, Dimensions, and Microsoft Academic for 18 selected Arab universities.

In Figure 2 we present a stacked bar for each database pair with the primary database on the left and the comparator database on the right. The first portion of the bar reading from the left represents the surplus records only found in the primary database and not present in the comparator. For instance, the top bar shows 30,060 Scopus DOIs that are not included in Web of Science. The second portion includes those DOIs that are present in both databases but are not assigned to the relevant university affiliation in the comparator database. In the first bar that means 1,831 Scopus DOIs are not found to have the same affiliation in Web of Science, even though they are present in Web of Science. The central portion of the bar shows the number of DOIs found in both databases with the same affiliation. The penultimate portion of the bar shows that 5,924 Web of Science DOIs are also found in Scopus but not with the relevant affiliation. In the final portion of the bar, there are 3,114 Web of Science DOIs that are not included in the Scopus database.

Those DOIs in the second and fourth portions of the bar therefore represent DOIs where there is a discrepancy between the affiliation assigned in the two databases. It is not possible at this point to say which, if either is wrong. To make a decision about the accuracy of the assigned affiliation, we need to look at the individual cases and usually check with the published PDF document. We report this in section 4.4 of the study. Therefore, we do not refer to affiliation errors, but prefer the term discrepancies. There will also be cases in which the affiliation is missing or has failed to be assigned to the university in both databases and those records would not appear at all in the results. We are therefore not presenting a comprehensive list of errors, but rather an indication of the proportion of discrepancies between databases pairs.

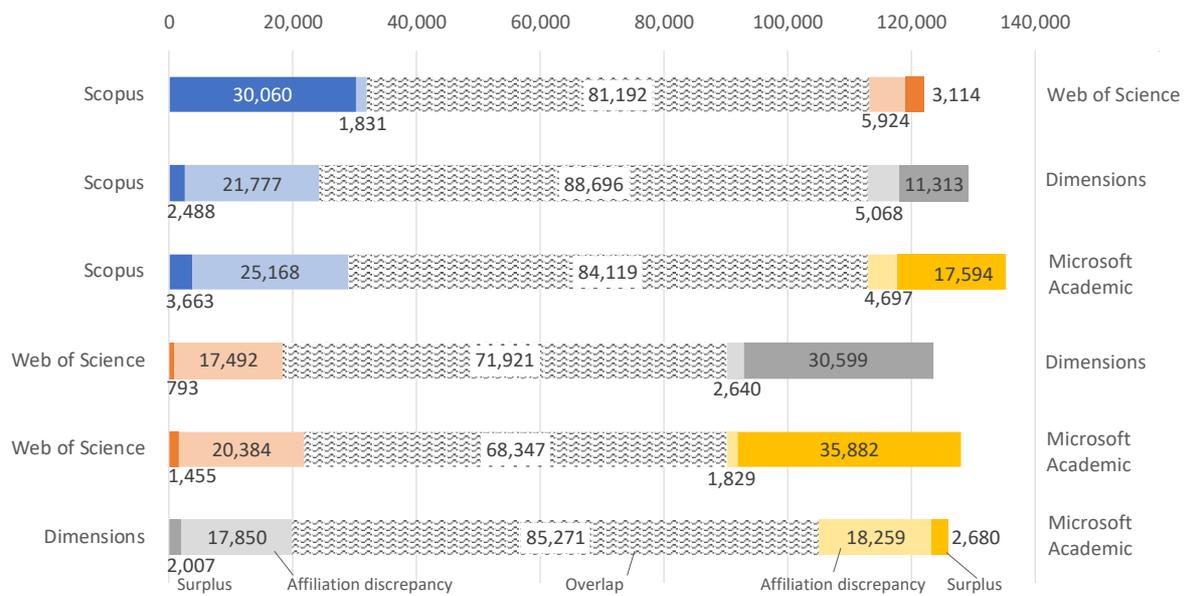

Figure 1. Affiliation discrepancy for the 18 universities by database pair

We can now analyse the relative affiliation discrepancies between each of the database pairs presented in Figure 3. For ease of discussion, we have organised the database pairs into four sections with the primary database as the heading. In each comparison, we only used those records associated with a DOI.

Scopus

The 1,831 Scopus records from the 18 selected universities that are present in Web of Science but not linked to the same affiliation represent about 2% of the actual overlap between the two databases. The next two bars show far more sizeable proportions of Scopus records not linked to the same affiliation in Dimensions and Microsoft Academic. We can interpret these results by suggesting the Scopus affiliation is more likely to agree with that assigned in Web of Science than it is with those assigned in either Dimensions or Microsoft Academic.

On the other hand, a small share (4% – 7%) of publications has been assigned to the 18 universities in Scopus, but not in Web of Science (5,924), Dimensions (5,068), and Microsoft Academic (4,697).

Web of Science

Records linked to the selected universities in Web of Science but not assigned to the same affiliations in each of the other three databases appeared to constitute about 7% of the overlapping Scopus records, and about a fifth of the overlapping DOIs in Dimensions (17,492) and Microsoft Academic (20,384). This means the Web of Science affiliation does not concur with the other three databases in a substantial proportion of cases.

Comparing in the other direction, the penultimate section of the Web of Science bars (1,831 records for Scopus, 2,640 for Dimensions, and 1,829 for Microsoft Academic) is relatively small in each case. That means there are comparatively few records in the three comparator databases in which the affiliations fail to concur with their corresponding record in Web of Science.

Dimensions

Dimensions records were clearly more likely to coincide with the affiliations assigned in Web of Science and Scopus than they were for affiliations assigned by Microsoft Academic. Dimensions affiliation discrepancies with those assigned in Web of Science (2,640 records) and Scopus (5,068 records) accounted for only 2% and 4% of the overlapping coverage respectively. Meanwhile, 15% Dimensions records with overlapping coverage in Microsoft Academic showed affiliation discrepancies with their corresponding records in Microsoft Academic.

The opposite comparison shows that a substantial number of publications have not been assigned to the 18 universities in Dimensions, while they have been assigned to these universities in the other three databases. This suggests that Dimensions may incorrectly not have assigned these publications to the 18 universities.

Microsoft Academic

Similarly, there was a relatively small number of Microsoft Academic records that did not agree with affiliations assigned in Web of Science (1,829) and Scopus (4,697). These discrepancies represented 2% and 4% of the overlapping records respectively. A more sizeable proportion of around 15% Microsoft Academic records failed to match the affiliations assigned in Dimensions.

From the other direction, a substantial number of publications have not been assigned to the 18 universities in Microsoft Academic, while they have been assigned to these universities in the other three databases. This suggests that Microsoft Academic may incorrectly not have assigned these publications to the 18 universities.

Overall summary of the results:

When Web of Science or Scopus assign an affiliation to a publication, the same affiliation is usually assigned by the other databases. The same is not the case the other way round. The largest share of discrepancies occurred with affiliations assigned by Microsoft Academic when compared with the other three databases. Dimensions assigned affiliations also found sizeable shares of discrepancies when their records were compared in the other databases. The database coverage played a role with many records from the larger databases (Dimensions and Microsoft Academic) not found indexed in the smaller, more selective databases (Scopus and especially Web of Science).

*4.3. Database surplus by university*

Figures 4 – 9 show the extent of differences in coverage and affiliation between databases for each of the 18 selected universities. The additional level of data can help us interpret the differences between the way the databases approach affiliation disambiguation.

There were some patterns that emerged from this analysis. For example, there was a consistently higher proportion of records assigned to the Lebanese University than for the other universities found in Scopus, Dimensions, and Microsoft Academic that were not assigned to this university in Web of Science. We should recall that for this study, we defined Web of Science records as belonging to the universities only if they were retrieved using the organization enhanced unification tool. If records contained the correct address but was not unified to its affiliation in the organization enhanced tool, then they would not have been retrieved.

Similarly, records assigned to Khalifa University in Scopus, Web of Science, and Dimensions were frequently not found assigned to that university in Microsoft Academic. Khalifa University is the result of a merger between three institutions that took place in 2017. Some of these records will be examined in Section 4.4 to determine how different databases approached the resulting affiliations.

In the case of AUB many publications have been assigned to the university in Web of Science and Microsoft Academic, while they have not been assigned to the university in Dimensions. As we will show in the next section, this was likely due to differing treatment of records from the American University of Beirut Medical Centre.

Babylon had a higher proportion of Scopus records than other universities that were not assigned to that university in Dimensions and Microsoft Academic. It should be noted that Scopus uses one affiliation identifier (AFID) for the main institution and assigns other AFIDs for the sub-units of the university based on the organisation hierarchy. However, this process has been conducted more vigorously for some institutions than it has for others, and that difference might influence the level of affiliation discrepancy found. For example, Babylon has only one Scopus AFID and the lowest or nearly lowest share of affiliation discrepancies in all the database pairings where Scopus is the comparator. Conversely, Carthage has one Scopus AFID for the main institution and 31 AFIDs for the sub-units which add 78% more records to the university total when they are all included in the search. As discussed in Section 3, we used only the main AFID to identify the publications of a university in Scopus. Hence, in the case of Carthage, the large number of AFIDs for sub-units of the university probably explains why the university has the highest or second highest share of affiliation discrepancies in all comparisons where Scopus was the comparator.

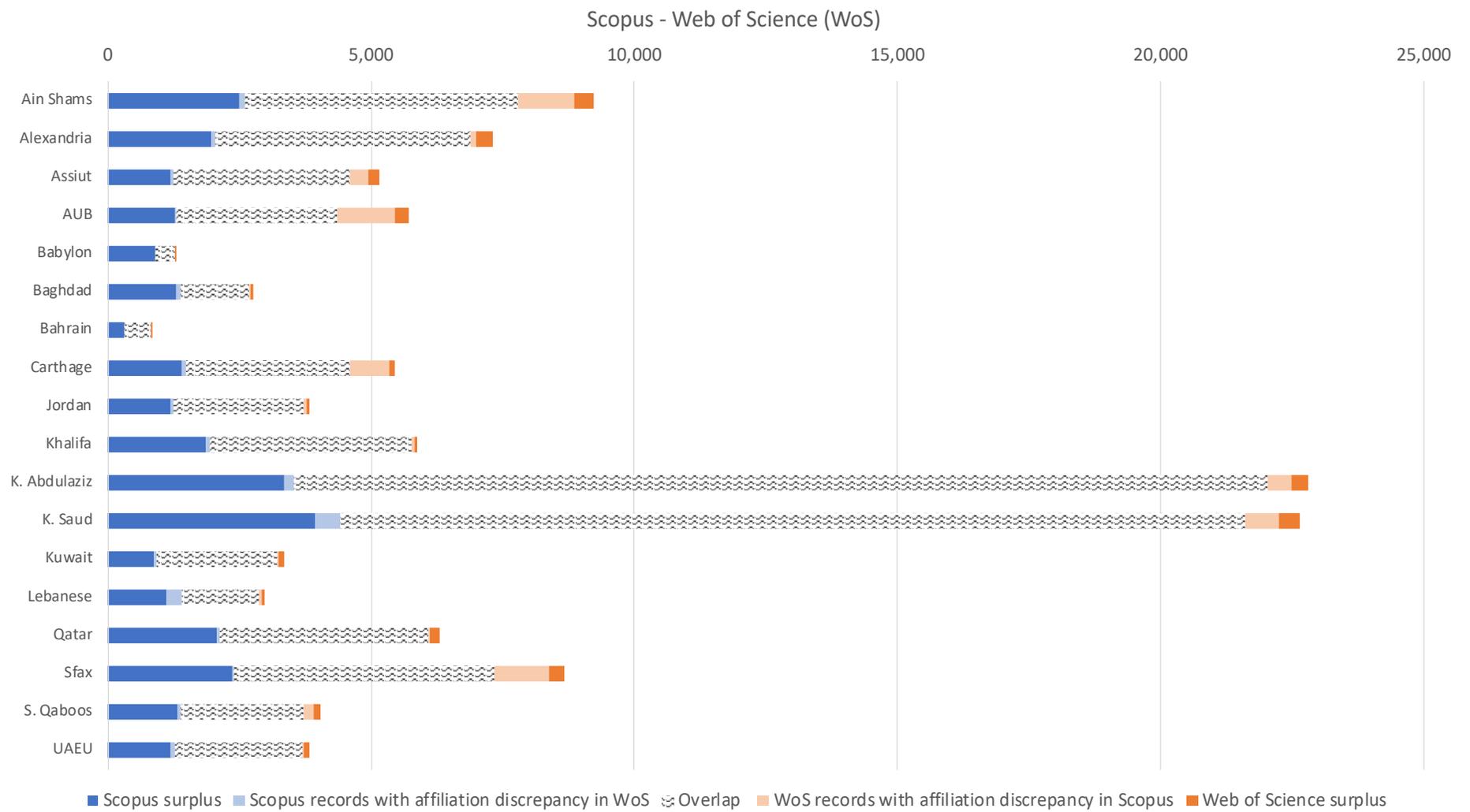

Figure 2. Scopus – Web of Science differences in coverage and affiliation by university

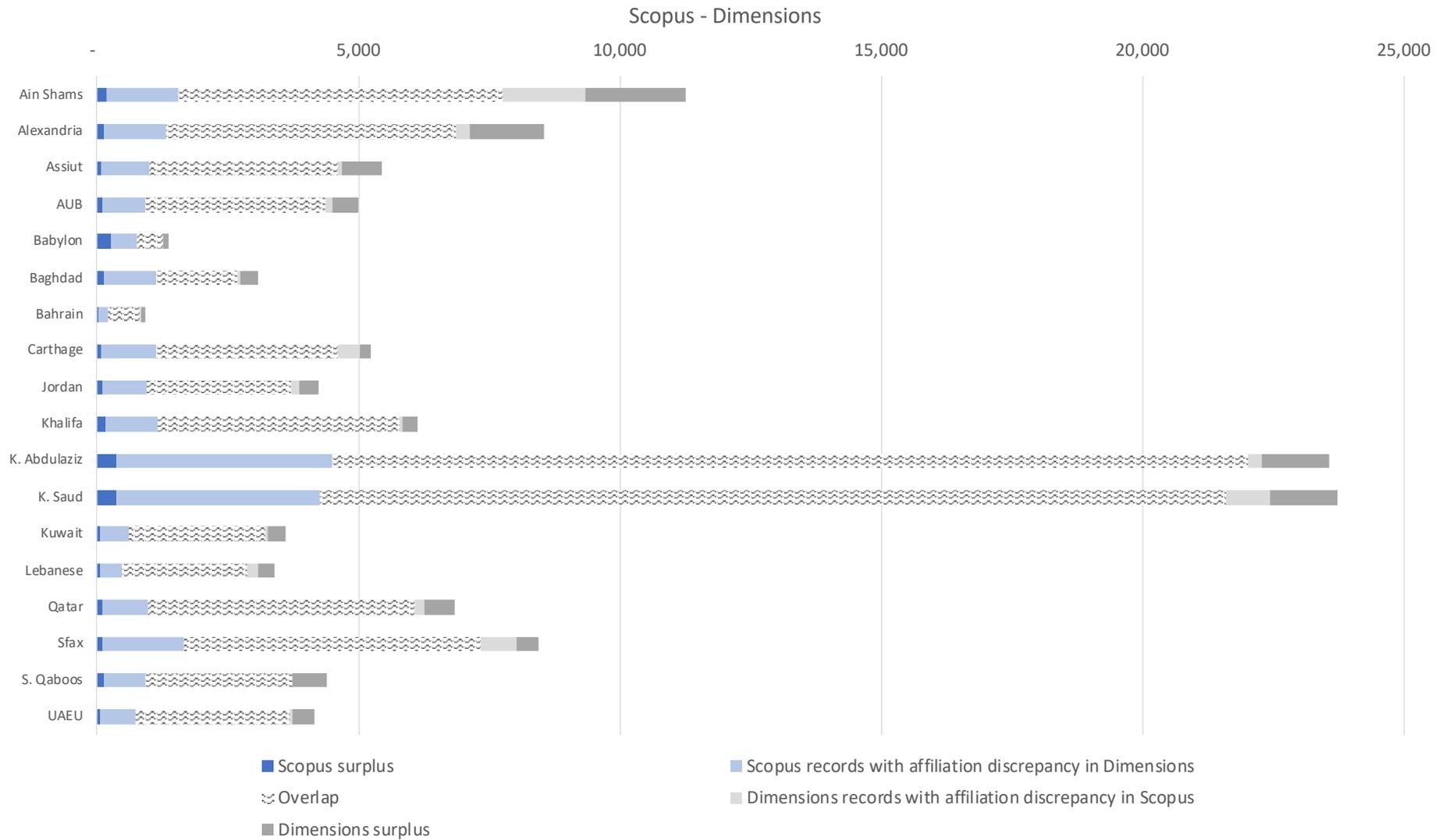

Figure 3. Scopus - Dimensions differences in coverage and affiliation by university

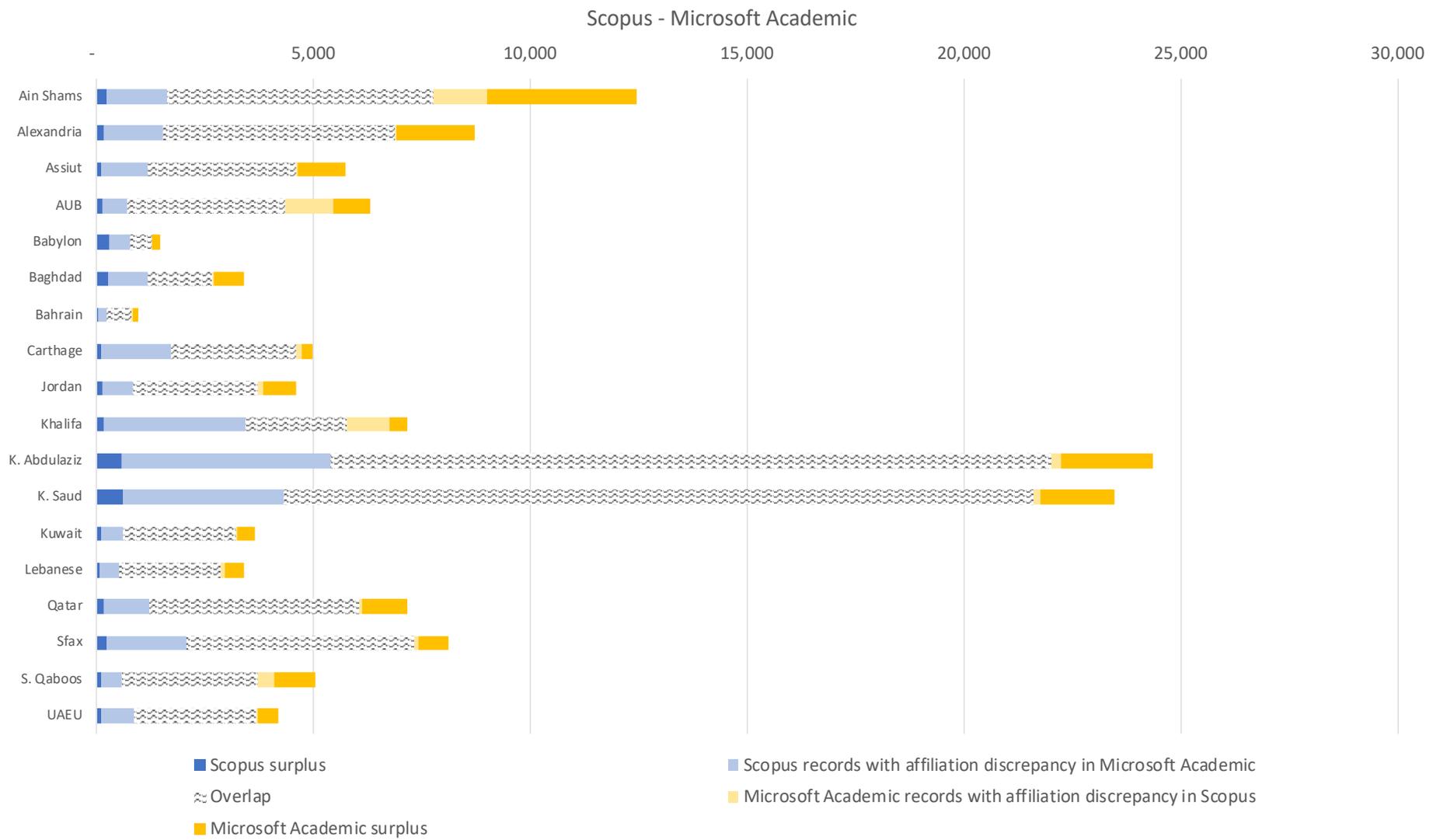

Figure 4. Scopus – Microsoft Academic differences in coverage and affiliation by university

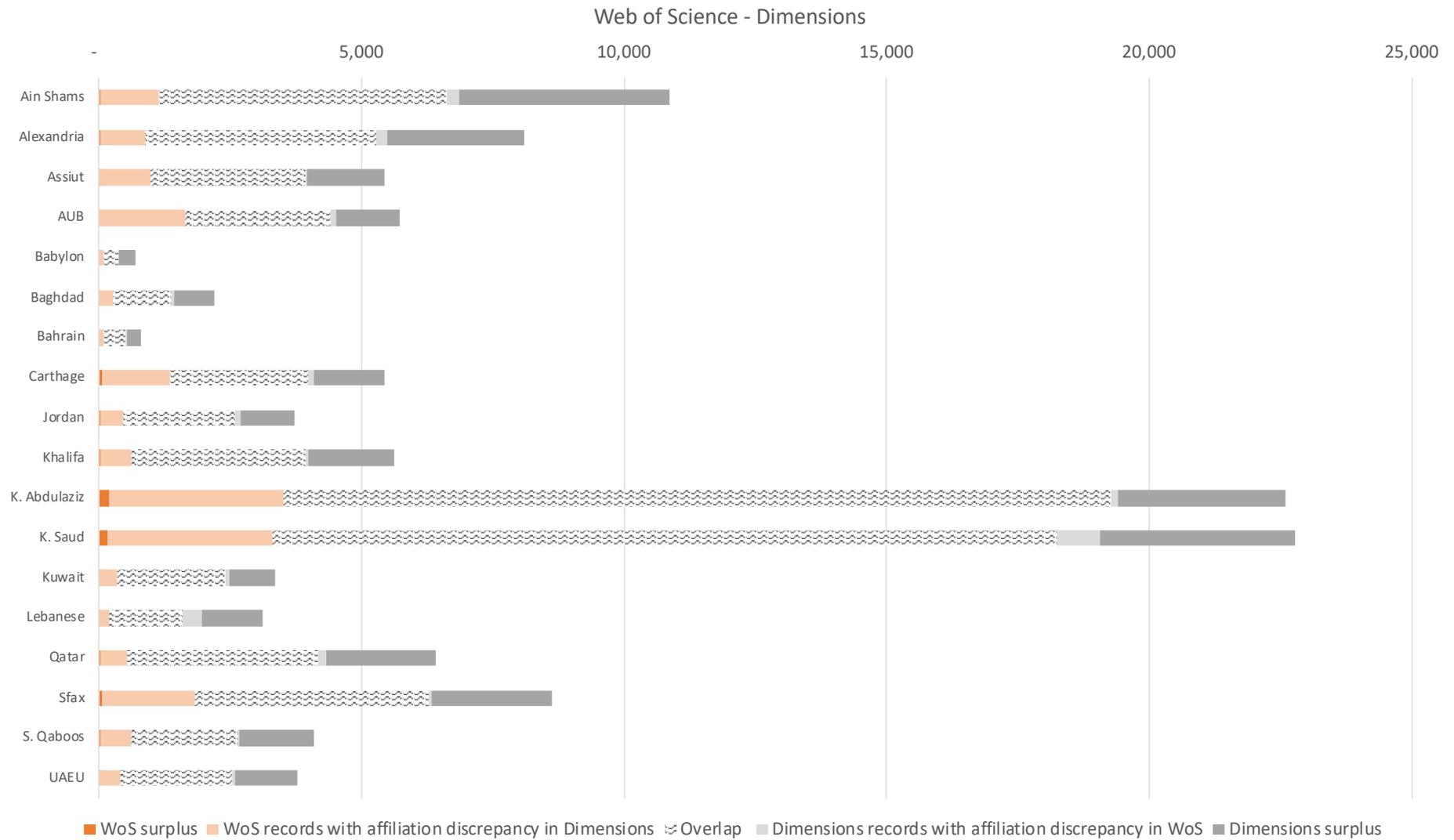

Figure 5. Web of Science – Dimensions differences in coverage and affiliation by university

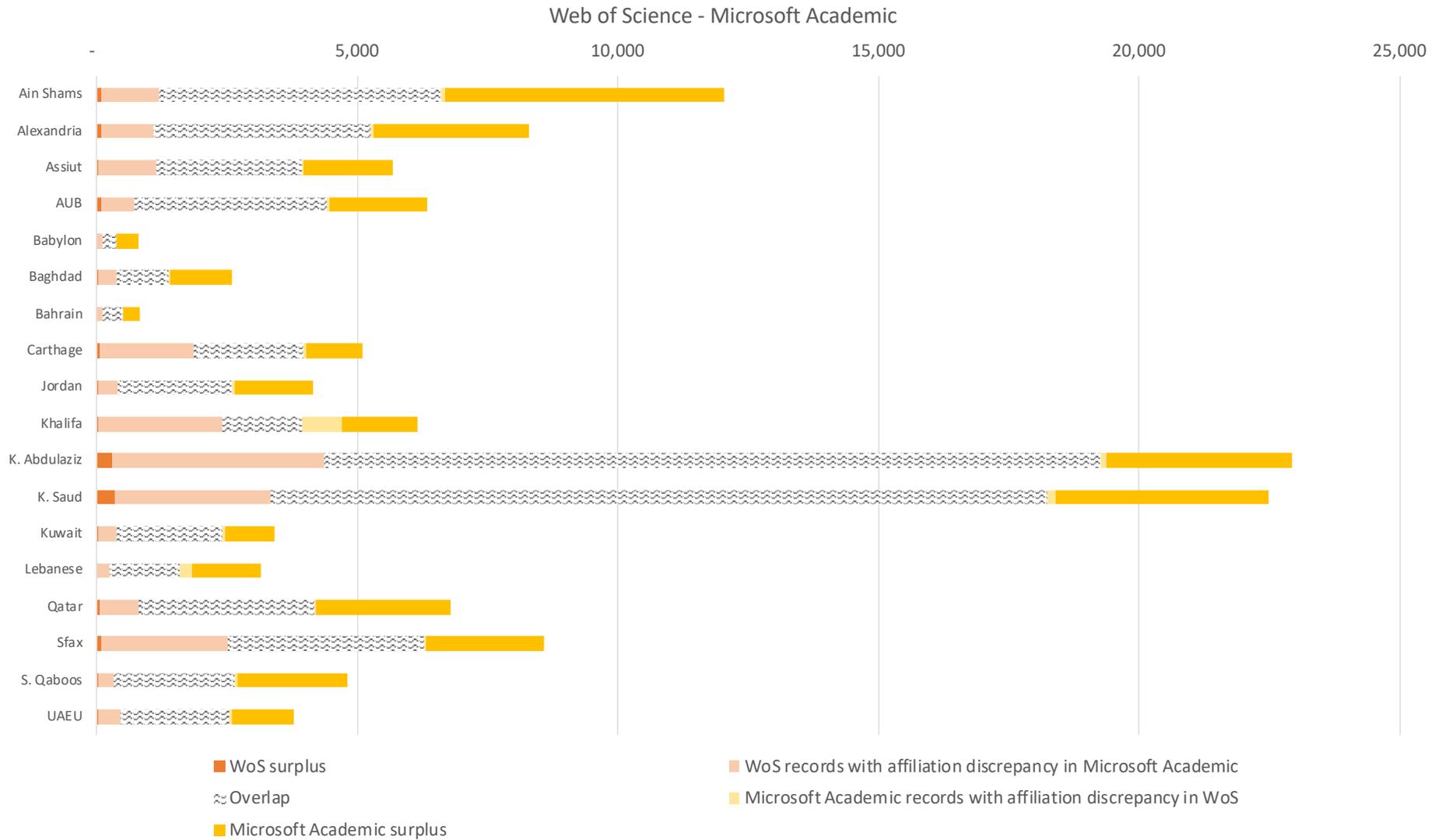

Figure 6. Web of Science – Microsoft Academic differences in coverage and affiliation by university

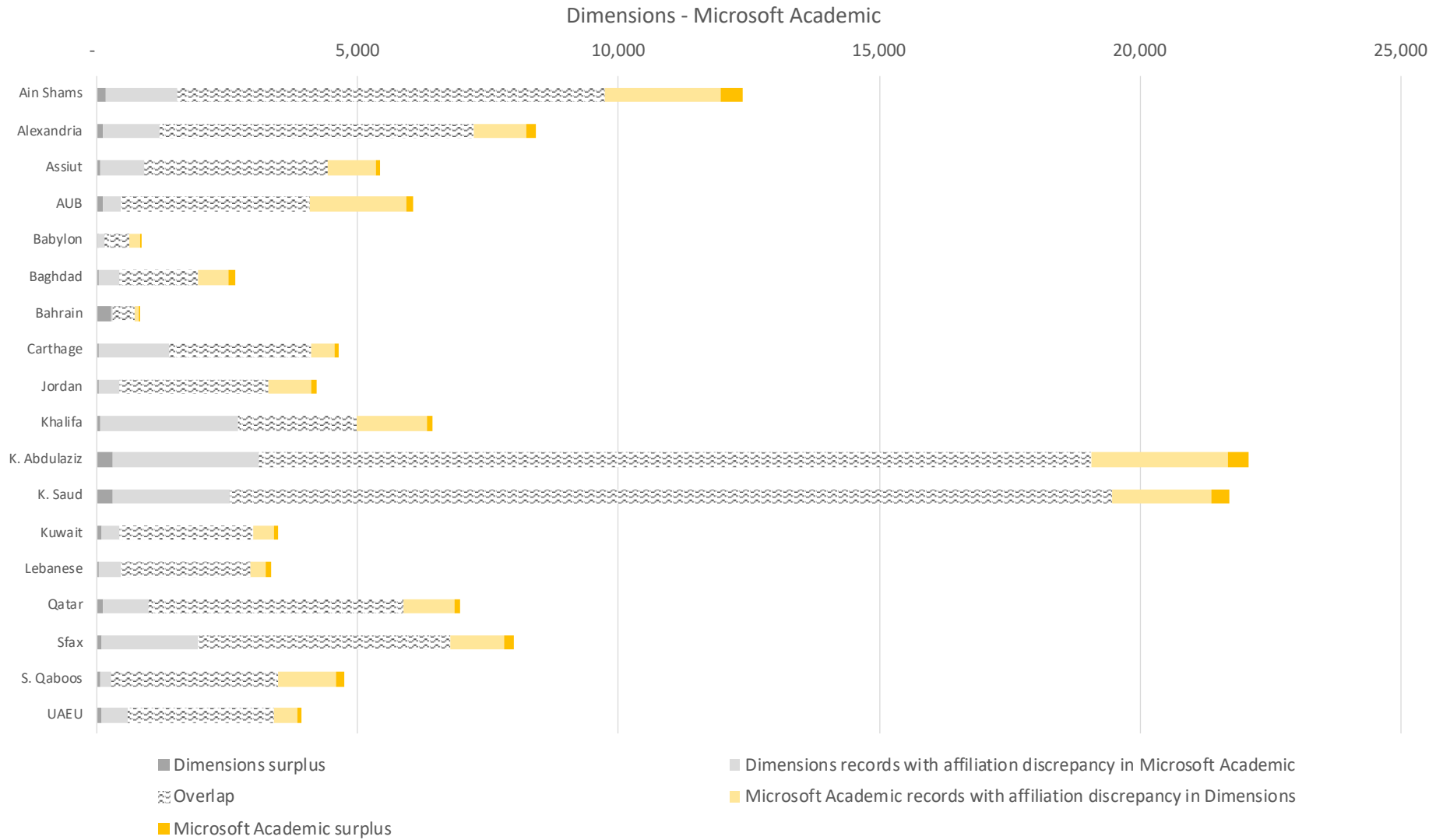

Figure 7. Dimensions – Microsoft Academic differences in coverage and affiliation by university

*4.4. Types of affiliation discrepancy*

The discrepancies in affiliations are the main focus of this study and vary widely in their prevalence and some interesting examples are described in this section. These highlight the challenges faced by database providers and the various ways they have responded to them. We found we could organise the affiliation discrepancies into four main groups as shown in Table 4.

Table 4. Types of affiliation discrepancy between databases

| Affiliation discrepancy type | Definition | Example | Database |
|---|---|---|---|
| Missing affiliation | Author's affiliation is missing | 10.4018/jdm.2016100102 | Scopus |
| Missing second affiliation | Author's first affiliation is present, but their second affiliation is missing | 10.1166/asl.2017.7424 | Dimensions |
| Unification | Affiliation mentioned in some form but not linked to unified record | 10.2174/1386207319666161214111822 | Web of Science |
| Assigned to wrong institution | Author address linked to a different institution than that intended | 10.1016/j.compfluid.2014.07.013 | Microsoft Academic |

We manually examined two dozen sample records at random from each of the database pairs using the web interface for each database. For each of these examples, we attempted to discover the reason for the affiliation discrepancy for one of the universities between the databases. The main reasons are summarised in Table 5. These sample analyses serve to illustrate that discrepancies exist and to shed light on the possible reasons behind them. However, a larger study would be needed to provide a more robust comparison.

Table 5. Affiliation errors

| Database pair | Institution | Missing affiliation | Missing second affiliation | Assigned to wrong institution | Unification | Inconclusive |
|---|---|---|---|---|---|---|
| Scopus - Web of Science | Lebanese | | | 1 | 22 | 1 |
| Scopus-Dimensions | Sfax | 11 | | | 11 | 2 |
| Scopus-Microsoft Academic | Khalifa | 2 | | | 20 | 2 |
| Web of Science - Scopus | Saud | 6 | 1 | 13 | | 4 |
| Web of Science - Dimensions | AUB | 6 | 1 | 1 | 14 | 2 |
| Web of Science - Microsoft Academic | Assiut | 6 | 12 | 5 | 1 | |
| Dimensions - Scopus | Carthage | 3 | | | 21 | |
| Dimensions - Web of Science | Bahrain | 1 | | 19 | 4 | |
| Dimensions - Microsoft Academic | Babylon | 9 | 6 | 7 | 2 | |
| Microsoft Academic - Scopus | Kuwait | | | | | 24 |
| Microsoft Academic - Web of Science | UAEU | 4 | 1 | 3 | 15 | 1 |
| Microsoft Academic - Dimensions | Qatar | 20 | 1 | 1 | 2 | |

### 4.4.1. *Missing affiliation*

The author affiliation has not been captured by the database and therefore a search for the affiliation name did not retrieve the record. We found almost all Qatar University records in the Microsoft Academic – Dimensions surplus were caused by missing affiliations in Dimensions. In most of these cases all author affiliations were missing, and the papers were conference proceedings or book chapters. Similarly, we found where Web of Science has missed author affiliations, the majority were meeting abstracts.

### 4.4.2. *Missing second affiliation*

The author's first affiliation has been listed but not the second. This is similar to the above category but worth separating as it appears that distinct groups of papers are indexed in some databases for which the additional affiliation is missed while the first is captured. As an example, on 10.1166/asl.2017.7424 the PDF shows University of Babylon as second affiliation for one author that is omitted from the record in Dimensions and Microsoft Academic, but included in Scopus and Web of Science. Sometimes we can speculate on the reason for this. In cases such as 10.1016/j.asoc.2016.06.019, the author's second affiliation is listed separately from the first on the PDF under categories such as 'Author's current address' or in this case, 'Correspondence address'. Scopus and Web of Science included this as second affiliation, while Dimensions and Microsoft Academic did not.

### 4.4.3. Assigned to the wrong institution

We found records from the College of Information Technology, University of Babylon and from the College of Information Technology, UAEU that had been erroneously assigned to the College of Information Technology, an institution registered in Pakistan in Microsoft Academic, e.g., 10.1007/s00500-018-3414-4. In other databases, these records had been correctly assigned to their respective universities. The term 'College of Information Technology' is a common university sub-unit, and it appears these words have triggered unification in Microsoft Academic to the stand-alone institution with the same name. Using these data in a bibliometric study would therefore produce a lower-than-expected count of papers from the affected institutions and an artificially high result for the real College of Information Technology in Pakistan. The same phenomenon occurs for Information Technology University in Pakistan, and the University College of Engineering in India, each of which are assigned additional papers in Microsoft Academic.

Other examples showed records assigned to UAEU in Web of Science organisation enhanced were in fact published by authors at a Moroccan institution called Université Abdelmalek Essaadi, locally abbreviated to 'UAE University' and mistakenly unified to the wrong institution. Similarly, authors from LaSTRe Laboratory in Tripoli, Northern Lebanon, which is affiliated to the Lebanese University, had been erroneously affiliated to the University of Benghazi in Tripoli, Libya in the Web of Science organisation enhanced. This might have occurred because of the appearance of the city name, Tripoli which is the capital of Libya but also a city in Lebanon. Two further records from the same university were assigned to the United States of America due to confusion over the town of Lebanon in Grafton County, New Hampshire. It appears therefore that the presence of a city name or country name might sometimes trigger unification to the wrong organisation enhanced name in the Web of Science.

### 4.4.4. Unification

An affiliation is listed but it has not been unified to the main or correct university. For example, most of the Scopus records that we did not find in Web of Science under the Lebanese University affiliation did actually mention the university in the address field but were not unified to that university in the organisation enhanced field. This is a plausible explanation for the large proportion of Lebanese university affiliation discrepancies with Web of Science described in section 4.3. Similarly, we discovered several examples of records attributed to either the Masdar Institute or Petroleum Institute in Abu Dhabi which have both been part of Khalifa University since a merger in 2017. Microsoft Academic has kept the original affiliation while the other databases have unified records to Khalifa University even from before the merger. This explains the high proportion of records unified to Khalifa in Scopus, Web of Science, and Dimensions but not found with that affiliation in Microsoft Academic. It also highlights the difficulties faced by database owners with treating records from organisations following their mergers or separations.

Many papers attributed to authors from the American University of Beirut Medical Centre have been unified to AUB in Web of Science, but not in Dimensions which treats it as a separate institution. That explains the notable proportion of affiliation discrepancies found with Dimensions in Figures 7 and 9. Similarly, many authors in Tunisia have acknowledged their institution as Faculty of Sciences at Sfax. We found that Scopus unified these papers to the University of Sfax while Dimensions treated it as a separate organisation. In another similar case also in Tunisia, we found several records affiliated to the National Institute of the Applied Sciences and Technology and several other research centres which Web of Science and Dimensions unified to the University of Carthage while Scopus did not. In most of these cases, the university was not mentioned on the PDF document, but Scopus has made the link through its disambiguation process. The Scopus interface offers users the option of searching the 'whole institutions' that includes all affiliated institutes or 'affiliation only' which is the main identifier for the university.

### 4.4.5. Inconclusive

We classified cases as inconclusive where there was no obvious reason for the DOI not being retrieved, a human indexing decision involved, or access to PDF proved impossible. An example of a human indexing decision is a book preface with a DOI (10.1016/B978-0-12-800887-4.00034-1) but no authors or affiliations. While Scopus had assigned the book editors as authors, Dimensions had not. Another interesting case was a letter to the editor published by three authors with a long list of additional signatories at the end. Scopus had counted all the signatories as authors of the letter while Dimensions limited authorship to the three at the top of the paper. Neither of these cases is clear cut, if one accepts the book editors in the first case, and letter signatories in the second should be named as authors then their affiliations are missing in Dimensions. If they should not be named, then they are phantom affiliations in Scopus.

### 4.5. Limitations

This study only used publications assigned a DOI in the comparisons between databases. As shown in Table 3 that excludes up to a fifth of the records depending on the database used which might also contain affiliation discrepancies and influence the results. In addition, as discussed in Section 2.1, there are potential errors in assigning DOIs including duplicate DOIs for the same publications, multiple publications, or versions of one publication with the same DOI, and various types of error within the DOI rendering it unable to link to its assigned publication. The DOI is the most appropriate identifier we found for use in this study, but we acknowledge its limitations.

The universities selected for use in the study were all from a specific geographical region and a broader comparison including institutions from different regions from around the world would show whether our findings generalise to other regions. Some databases have engaged with the institutions in the study to varying degrees as evidenced by the range in the number of affiliation variants listed among the Scopus AFIDs and Web of Science organization enhanced names. The engagement with universities will clearly improve the ability of the database to accurately identify and assign publications and therefore introduce a variable in our results.

The examination of types of affiliation discrepancy summarised in Table 5 relied on manual work and introduced an element of human judgement when comparing the published PDF document with its corresponding record in the bibliographic databases. This combined with the limited sample size for each comparison requires the reader to interpret the results as illustrations of the type of discrepancy and give an idea of the size of the problem. However, we do not interpret these data as statistically representative of the full, global databases.

## 5. Conclusions

Our results showed the proportion of Scopus and Web of Science records with DOIs was in line with previous studies and that Dimensions showed near universal DOI coverage. Microsoft Academic included a large proportion of content including patents and other non-academic document types that do not have DOIs. There was some variance in DOI coverage between universities probably due to prevalence of certain document types or subject fields which vary in their DOI assignation. Overall, bibliometric studies like the one presented in this paper can use the presence of DOIs to limit their datasets to comparable scholarly material.

We analysed overlapping coverage between databases in pairs and organised the results based on whether or not the author affiliations matched. Some records were assigned to the same university in both databases in the paired comparison. A substantial share of records were assigned to a university in one database but not in the other. This study concentrated on the affiliation discrepancies between the databases for 18 selected Arab universities.

We found evidence that up to one in five publications can have discrepancies in author affiliations between the major bibliographic databases. We found the discrepancies more frequently in the larger databases, Dimensions and Microsoft Academic. The highest incidences of discrepancy were Web of Science records not found to have the same affiliation in Microsoft Academic and then Dimensions. The next highest were Scopus records not found to have the same affiliation in Microsoft Academic and then Dimensions. Meanwhile, when publications were assigned affiliations in any of the databases, the same affiliation was usually assigned in Scopus, and especially in Web of Science. These two databases are smaller, more selective, and crucially, frequently engage with institutions to improve the unification of affiliation variants. Our Web of Science results might be more favourable than the real situation because our version excludes book chapters and a lot of geographically diverse journals often from university presses that might less rigorously assign author affiliations.

Our results revealed different reasons behind the discrepancies. These included problems of unifying address variants to the main institution, publications with missing author affiliations, and cases of records clearly being assigned to the wrong institution. One common source of difficulty was the naming of institutions like the College of Information Technology which is also the name of the sub-unit of many universities around the world. Another arises from cities of the same name in different countries such as Tripoli that exists in both Libya and Lebanon.

In our sample, we found that discrepancies in Web of Science were most frequently due to problems with unifying variants and in some cases, confusion clearly led to assigning records to the wrong institution. Discrepancies among Dimensions records were more likely found due to missing affiliations but there were also some issues with unification. In Scopus and Microsoft Academic there was no clear pattern and causes of discrepancy were mixed.

The Scopus affiliation identifier (AFID) and the Web of Science organisation enhanced feature each depend on engagement with the institutions to link affiliation variants to the main organisation. Where Scopus has assigned multiple AFIDs to sub-units of universities, such as the 31 AFIDs assigned to sub-units of the University of Carthage there is more chance of discrepancy with other databases. When comparing institutions, bibliometricians should resist the assumption the disambiguation process has been performed to the same level for all institutions in the analysis. Our results show this is not always the case and that some comparisons will produce misleading results.

Manual examination of individual records revealed examples of publication records where databases have made different choices about how to unify author affiliations. The correct answer is often not clear and using one or another database will incorporate the impact of the choices of database owners into the results of any bibliometric analysis or benchmarking exercise based upon them.

University rankings providers usually rely on the disambiguation used in the source databases. The major exception is the Leiden Ranking which disambiguates all affiliations from its proprietary database (Calero-Medina et al., 2020), while both QS (QSIU, 2019) and Times Higher Education have begun supplementary work on Scopus unification in special cases. This practice is welcomed and providers of products that derive from bibliometric data sources should be encouraged to increasingly participate in the analytical process and assume a share of responsibility for the accuracy of the resulting publication.

These results support the conclusions of Huang et al. (2020) who encourages university ranking publishers to employ multiple bibliographic data sources. While Visser et al. (2021) discuss the merits of more selective databases for university rankings, the results in this paper show the presence of author affiliation disambiguation in Scopus and Web of Science still poses a significant limitation to accuracy.

A considerable limitation to the present and most previous studies on affiliation disambiguation is the fact that university names change over time. Changes result from a number of factors including mergers and splits but also for reasons of government naming conventions, changes of country leaders, city names, and other factors. Once an organisational sub-unit is unified to an affiliation, all its prior papers will be found when searching the new unified affiliation. Studies are therefore a time-frozen shot of the current unification and do not take account of unification dynamics over time.

There is a need for a universal unique identifier for academic institutions that should reflect the current and historical organisation relationship tree. The ideal indicator will be supported by input from the institutions themselves in the same way that researchers maintain their own ORCID records. That way the accuracy, maintenance, and historical record will be maximised. There will be scope for non-maintenance or misuse especially where an institution can benefit from a certain interpretation of its organisation, but these will be outweighed by the benefits. In addition, in the case of an open infrastructure, any misuse will be publicly visible which will act as a disincentive. Universities and their stakeholders should still decide their own names and they are still the most appropriate managers of the public record of their relationships with sub-units and external entities.

This study has demonstrated the scope for improvement in four bibliographic databases and highlighted many problems faced by those attempting the task of disambiguation. Many of those cases can potentially be resolved by incorporating a global, universally accepted identifier for use by the worldwide research community and supported with input from universities.

## Acknowledgements

Ton van Raan and Ludo Waltman for expert guidance throughout the study, Christian Herzog (Digital Science) and Martin Szomszor (Clarivate) for their useful comments on earlier versions of the article, two anonymous reviewers whose suggestions helped improve the manuscript.

Table 6. Data table from Figure 2. Scopus – Web of Science affiliation discrepancy by university

| | Scopus surplus | Scopus records with affiliation discrepancy in WoS | Overlap | WoS records with affiliation discrepancy in Scopus | Web of Science surplus |
|---|---|---|---|---|---|
| Ain Shams | 2,506 | 92 | 5,187 | 1,077 | 357 |
| Alexandria | 1,956 | 90 | 4,829 | 106 | 342 |
| Assiut | 1,185 | 56 | 3,367 | 345 | 218 |
| AUB | 1,263 | 39 | 3,062 | 1,080 | 278 |
| Babylon | 891 | 4 | 374 | 3 | 12 |
| Baghdad | 1,288 | 97 | 1,302 | 16 | 61 |
| Bahrain | 313 | 11 | 487 | 4 | 15 |
| Carthage | 1,398 | 82 | 3,115 | 757 | 110 |
| Jordan | 1,197 | 42 | 2,486 | 50 | 63 |
| Khalifa | 1,869 | 80 | 3,832 | 46 | 58 |
| K. Abdulaziz | 3,345 | 185 | 18,507 | 445 | 334 |
| K. Saud | 3,931 | 478 | 17,210 | 620 | 415 |
| Kuwait | 872 | 65 | 2,267 | 40 | 99 |
| Lebanese | 1,113 | 284 | 1,482 | 39 | 69 |
| Qatar | 2,067 | 45 | 3,968 | 45 | 169 |
| Sfax | 2,353 | 39 | 4,947 | 1,036 | 309 |
| S. Qaboos | 1,320 | 58 | 2,348 | 192 | 110 |
| UAEU | 1,193 | 84 | 2,422 | 23 | 95 |
| **Totals** | **30,060** | **1,831** | **81,192** | **5,924** | **3,114** |

Table 7. Data table from Figure 3. Scopus – Dimensions affiliation discrepancy by university

| | Scopus surplus | Scopus records with affiliation discrepancy in Dimensions | Overlap | Dimensions records with affiliation discrepancy in Scopus | Dimensions surplus |
|---|---|---|---|---|---|
| Ain Shams | 182 | 1,361 | 6,228 | 1,565 | 1,933 |
| Alexandria | 125 | 1,199 | 5,544 | 250 | 1,423 |
| Assiut | 77 | 921 | 3,605 | 75 | 755 |
| AUB | 96 | 836 | 3,425 | 153 | 501 |
| Babylon | 254 | 505 | 506 | 2 | 110 |
| Baghdad | 126 | 1,016 | 1,540 | 61 | 329 |
| Bahrain | 38 | 160 | 612 | 27 | 83 |
| Carthage | 86 | 1,052 | 3,453 | 424 | 216 |
| Jordan | 93 | 866 | 2,759 | 157 | 356 |
| Khalifa | 151 | 997 | 4,628 | 59 | 304 |
| K. Abdulaziz | 377 | 4,113 | 17,518 | 274 | 1,286 |
| K. Saud | 380 | 3,877 | 17,339 | 844 | 1,283 |

| | | | | |
|---|---|---|---|---|
| Kuwait | 62 | 545 | 2,596 | 68 | 337 |
| Lebanese | 54 | 419 | 2,406 | 192 | 334 |
| Qatar | 100 | 884 | 5,094 | 178 | 572 |
| Sfax | 103 | 1,551 | 5,679 | 686 | 423 |
| S. Qaboos | 127 | 787 | 2,813 | 21 | 651 |
| UAEU | 57 | 688 | 2,951 | 32 | 417 |
| **Totals** | **2,488** | **21,777** | **88,696** | **5,068** | **11,313** |

Table 8. Data table from Figure 4. Scopus – Microsoft Academic affiliation discrepancy by university

| | Scopus surplus | Scopus records with affiliation discrepancy in Microsoft Academic | Overlap | Microsoft Academic records with affiliation discrepancy in Scopus | Microsoft Academic surplus |
|---|---|---|---|---|---|
| Ain Shams | 240 | 1,388 | 6,140 | 1,232 | 3,450 |
| Alexandria | 183 | 1,347 | 5,338 | 49 | 1,811 |
| Assiut | 110 | 1,084 | 3,411 | 34 | 1,093 |
| AUB | 154 | 552 | 3,656 | 1,094 | 855 |
| Babylon | 283 | 497 | 485 | 3 | 213 |
| Baghdad | 260 | 915 | 1,510 | 14 | 690 |
| Bahrain | 31 | 209 | 569 | 11 | 142 |
| Carthage | 107 | 1,620 | 2,862 | 132 | 260 |
| Jordan | 131 | 713 | 2,871 | 116 | 764 |
| Khalifa | 159 | 3,276 | 2,336 | 989 | 411 |
| K. Abdulaziz | 571 | 4,836 | 16,604 | 233 | 2,123 |
| K. Saud | 629 | 3,676 | 17,299 | 163 | 1,702 |

| | | | | |
|---|---|---|---|---|
| Kuwait | 116 | 513 | 2,573 | 40 | 426 |
| Lebanese | 66 | 441 | 2,369 | 71 | 443 |
| Qatar | 186 | 1,015 | 4,871 | 48 | 1,061 |
| Sfax | 221 | 1,850 | 5,257 | 85 | 712 |
| S. Qaboos | 107 | 466 | 3,151 | 357 | 955 |
| UAEU | 109 | 770 | 2,817 | 26 | 483 |
| **Totals** | **3,663** | **25,168** | **84,119** | **4,697** | **17,594** |

Table 9. Data table from Figure 5. Web of Science – Dimensions affiliation discrepancy by university

| | WoS surplus | WoS records with affiliation discrepancy in Dimensions | Overlap | Dimensions records with affiliation discrepancy in WoS | Dimensions surplus |
|---|---|---|---|---|---|
| Ain Shams | 43 | 1,088 | 5,487 | 244 | 4,010 |
| Alexandria | 49 | 842 | 4,383 | 215 | 2,620 |
| Assiut | 19 | 975 | 2,935 | 47 | 1,457 |
| AUB | 22 | 1,630 | 2,767 | 100 | 1,213 |
| Babylon | 6 | 72 | 311 | 2 | 305 |
| Baghdad | 11 | 264 | 1,103 | 58 | 769 |
| Bahrain | 6 | 71 | 428 | 31 | 263 |
| Carthage | 53 | 1,291 | 2,638 | 102 | 1,363 |
| Jordan | 39 | 416 | 2,141 | 114 | 1,023 |
| Khalifa | 30 | 593 | 3,313 | 53 | 1,625 |
| K. Abdulaziz | 182 | 3,338 | 15,760 | 125 | 3,195 |
| K. Saud | 166 | 3,150 | 14,924 | 815 | 3,729 |
| Kuwait | 23 | 331 | 2,052 | 72 | 877 |

| | | | | |
|---|---|---|---|---|
| Lebanese | 17 | 176 | 1,397 | 367 | 1,168 |
| Qatar | 34 | 503 | 3,645 | 161 | 2,082 |
| Sfax | 53 | 1,783 | 4,458 | 44 | 2,291 |
| S. Qaboos | 25 | 586 | 2,038 | 35 | 1,412 |
| UAEU | 15 | 383 | 2,141 | 55 | 1,197 |
| **Totals** | **793** | **17,492** | **71,921** | **2,640** | **30,599** |

Table 10. Data table from Figure 6. Web of Science – Microsoft Academic affiliation discrepancy by university

| | WoS surplus | WoS records with affiliation discrepancy in Microsoft Academic | Overlap | Microsoft Academic records with affiliation discrepancy in WoS | Microsoft Academic surplus |
|---|---|---|---|---|---|
| Ain Shams | 91 | 1,106 | 5,418 | 65 | 5,351 |
| Alexandria | 86 | 1,012 | 4,174 | 46 | 2,978 |
| Assiut | 50 | 1,082 | 2,797 | 26 | 1,715 |
| AUB | 89 | 638 | 3,693 | 61 | 1,866 |
| Babylon | 11 | 93 | 285 | 2 | 414 |
| Baghdad | 37 | 340 | 1,002 | 19 | 1,193 |
| Bahrain | 5 | 101 | 400 | 14 | 308 |
| Carthage | 63 | 1,787 | 2,130 | 42 | 1,085 |
| Jordan | 50 | 344 | 2,202 | 61 | 1,495 |
| Khalifa | 43 | 2,370 | 1,520 | 774 | 1,450 |
| K. Abdulaziz | 290 | 4,077 | 14,912 | 105 | 3,538 |
| K. Saud | 356 | 2,968 | 14,913 | 164 | 4,086 |
| Kuwait | 36 | 348 | 2,022 | 51 | 966 |
| Lebanese | 23 | 232 | 1,334 | 247 | 1,302 |
| Qatar | 56 | 747 | 3,378 | 26 | 2,576 |
| Sfax | 94 | 2,427 | 3,767 | 22 | 2,266 |
| S. Qaboos | 35 | 302 | 2,311 | 49 | 2,110 |
| UAEU | 40 | 410 | 2,089 | 55 | 1,183 |
| **Totals** | **1,455** | **20,384** | **68,347** | **1,829** | **35,882** |

Table 11. Data table from Figure 7. Dimensions – Microsoft Academic affiliation discrepancy by university

| | Dimensions surplus | Dimensions records with affiliation discrepancy in Microsoft Academic | Overlap | Microsoft Academic records with affiliation discrepancy in Dimensions | Microsoft Academic surplus |
|---|---|---|---|---|---|
| Ain Shams | 179 | 1,377 | 8,185 | 2,228 | 418 |
| Alexandria | 113 | 1,096 | 6,007 | 1,018 | 171 |
| Assiut | 55 | 851 | 3,530 | 904 | 101 |
| AUB | 120 | 329 | 3,631 | 1,858 | 120 |
| Babylon | 5 | 148 | 465 | 201 | 35 |
| Baghdad | 37 | 396 | 1,497 | 587 | 130 |
| Bahrain | 263 | 31 | 428 | 71 | 6 |
| Carthage | 46 | 1,339 | 2,715 | 453 | 87 |
| Jordan | 48 | 398 | 2,830 | 828 | 99 |
| Khalifa | 76 | 2,624 | 2,289 | 1,343 | 109 |
| K. Abdulaziz | 294 | 2,817 | 15,961 | 2,606 | 392 |
| K. Saud | 301 | 2,257 | 16,910 | 1,896 | 357 |
| Kuwait | 87 | 344 | 2,569 | 391 | 80 |
| Lebanese | 40 | 408 | 2,484 | 315 | 85 |
| Qatar | 111 | 871 | 4,901 | 969 | 108 |
| Sfax | 77 | 1,854 | 4,858 | 1,029 | 164 |
| S. Qaboos | 74 | 193 | 3,218 | 1,098 | 153 |
| UAEU | 81 | 517 | 2,793 | 464 | 65 |
| **Totals** | **2,007** | **17,850** | **85,271** | **18,259** | **2,680** |